\documentclass[pra,twocolumn,preprintnumbers,amsmath,amssymb,showpacs,
nofootinbib,floatfix]{revtex4}

\usepackage{graphicx,bm,amsmath}
\usepackage{bm}
\usepackage{tikz}

\makeatletter

\def\graphicscale{\twocolumn@sw{0.30}{0.33}}
\def\graphicthreescale{\twocolumn@sw{0.30}{0.33}}

\makeatother

\begin{document}

\title{Scaling behaviour of trapped bosonic particles in two dimensions at finite temperature}

 \author{Giacomo Ceccarelli, Christian Torrero}
   \affiliation{Dipartimento di Fisica dell'Universit\`a di Pisa and
   I.N.F.N., Sezione di Pisa, Largo Bruno Pontecorvo 2, I-56127 Pisa,
   Italy} \date{May 7, 2012}

\begin{abstract}
In the framework of the trap-size scaling theory, we study 
the scaling properties of the Bose-Hubbard model in two dimensions 
in the presence of a trapping potential at finite temperature. 
In particular, we provide results for the particle density and 
the density-density correlator at the Mott transitions and within 
the superfluid phase. For the former quantity, numerical outcomes are also extensively compared to 
Local Density Approximation predictions. 
\end{abstract}

% Ultracold gases trapped gases, Quantum phase transitions, Boson systems, 

\maketitle

% ========================= BODY =========================
%\narrowtext

\section{Introduction}
\label{intro}
Recently, optical lattices have acquired primary experimental importance since they allow the study of the main features of systems of well-localized cold atoms: among others, the intriguing interplay of thermal and quantum effects in bosonic gases can thus be investigated rather precisely, with particular attention usually paid to Mott-Hubbard transitions~\cite{BDZ-08,GBMHS-02,HSBBD-06,SPP-07,FWMGB-06}. 
A peculiarity of such experiments is the confining of particles within a limited region of the lattice which is normally achieved by introducing a trapping potential. This experimental setup can be mimicked theoretically by the so-called Bose-Hubbard (BH) Hamiltonian~\cite{FWGF-89} reading
\begin{eqnarray}
H_{\rm BH} &=& -{J\over 2}
\sum_{\langle ij\rangle}(b_i^\dagger b_j+b_j^\dagger b_i)
+ {U\over2} \sum_i n_i(n_i-1)+
\nonumber \\&&+
\mu \sum_i n_i + \sum_i V(r_i) n_i ,
\label{bhm}
\end{eqnarray}    
where $b^\dagger_i$ and $b_i$ are respectively bosonic creation and destruction operators, $n_i$ is the local density
 operator, $\mu$ is the chemical potential, $U$ is the on-site repulsion energy and the sum in the first term
is over the nearest-neighbor sites of a regular $d$-dimensional lattice. As for the trapping potential $V(r_i)$ (being $r_i$ the distance from the center of the trap), a common choice is given by
\begin{equation}
V(r_i)= v^p r_i^p,
\label{vrp}
\end{equation}
being $l \equiv J^{1/p}v^{-1}$ the trap size. The exponent $p$ is clearly even and will be set to $p=2$ in the following. Moreover, the energy unit will be fixed by setting $J=1$ (thus $l=1/v$) while $r$ and $l$ will be measured in units of the lattice spacing $a$~\!\footnote{~\!$a$ will be set to $1$ from now on.} and hence dimensionless.

In the homogenous case (i.e., with vanishing trap) the model undergoes quantum transitions between superfluid and Mott-insulator phases depending on the value of $\mu$.% in such a critical regime, the correlation length $\xi$ diverges with critical exponent $\nu=1/2$ while the dynamic exponent $z$ equals to $z=2$ \cite{FWGF-89}.

\noindent The introduction of a trapping potential changes the phase diagram~\cite{BRSRMDT-02,WATB-04,GKTWB-06,KPS-02,KSDZ-04}: not only a truly diverging correlation length appears only in the limit $l\rightarrow+\infty$~\cite{BRSRMDT-02,WATB-04} with $\mu$ set to the critical values of the corresponding homogeneous system, but also the scaling properties of any observable generally acquire an extra dependence on the trap size $l$ controlled by the trap exponent $\theta$ given by
\begin{equation}
\theta= {p\over p+2}.
\label{thetaexp}
\end{equation}
A frame to handle this involved scaling is provided by the trap-size scaling (TSS) theory~\cite{CV-10,CV-09}. As a benchmark example, at a quantum critical point TSS prescribes the free-energy density to scale as 
\begin{equation}
F(\mu,T,l,r) = l^{-\theta(d+z)} 
{\cal F}(\bar{\mu} l^{\theta/\nu},Tl^{\theta z},rl^{-\theta}),
\label{freee}
\end{equation}
with $z$ the dynamical exponent, $\nu$ the critical exponent controlling how the correlation length diverges, $r$ the distance from the middle of the trap,
$\bar{\mu}\equiv \mu-\mu_c$, and $\mu_c$ the critical value of the chemical potential.

TSS has already been applied to the one-dimensional (1D) BH model, both at $T=0$~\cite{CV-10-bh} and at finite temperature~\cite{CTV}: in this paper we extend it to the two-dimensional (2D) BH model at finite temperature. Indeed, 2D systems are relevant not only from a theoretical point of view but have also raised experimental interest~\cite{SPP-07,JCLPPS-10,SPP-08}.

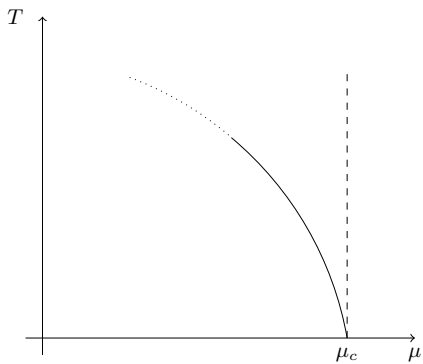
\begin{figure}
\begin{center}
\scalebox{0.9}
{
    \begin{tikzpicture}
        \draw[->] (0.75,1)--(6.5,1);
        \draw[->] (1,0.75)--(1,5.75);
        \draw (0.6,5.75) node {$T$};
        \draw (6.5,0.75) node {$\mu$};
        \draw (5.5,0.75) node {$\mu_c$};
        \draw (5.5cm,1cm) arc (10:50:5cm);
        \draw[dashed] (5.5,1)--(5.5,5);
        \draw[dotted] (3.75cm,4cm) arc (50:70:5cm);
    \end{tikzpicture}
}
\end{center}
\caption{A qualitative sketch of the Kosterlitz-Thouless transition (solid line) in the $\mu-T$ plane.}
\label{KT}
\end{figure}

At $T=0$, the 2D BH model (\ref{bhm}) in the hard-core limit (see below) undergoes two phase transitions between superfluid and Mott insulator at $\mu=2$ and $\mu=-2$~\!\footnote{~\!More precisely, the system is in a Mott phase with $\langle n_i\rangle=0$ for $\mu>2$, in a superfluid phase for $|\mu|<2$ and in a Mott phase with $\langle n_i\rangle=1$ for $\mu<-2$. The two transitions share the same critical exponents $\nu=1/2$ and $z=$2 \cite{FWGF-89}.}, while at finite temperature it is well known that the model also develops a Kosterlitz-Thouless (KT) transition~\cite{KT-73,B-72}. In this work we are not going to study the latter but rather perform quantum Monte Carlo (QMC) simulations with chemical potential fixed at $\mu=-2, 0, 2$ only while varying $T$: This is because our aim is to investigate the behavior of the model at the quantum $T=0$ critical points. This choice of the parameters should avoid any crossings of the KT line, as clear from the qualitative diagram in Fig. \ref{KT}.  

In this framework, the scaling of the particle density 
\begin{equation}
\rho(r_i) \equiv \langle n_{i} \rangle\ ,
\end{equation}
and the density-density correlator 
\begin{equation}
G(r_i,r_j)  \equiv \langle n_{i} n_{j} \rangle - \langle n_{i} \rangle \langle n_{j} \rangle\ ,
\end{equation}
will be studied at fixed trap size $l$ and compared with TSS predictions. Besides, we extensively study how numerical outcomes for the particle density approach their Local Density Approximation (LDA) predictions at the Mott-to-superfluid transition with non-zero filling and within the superfluid phase.\\
\indent Since scaling properties are expected to be universal with respect to $U$, we will work in the hard-core (HC) limit $U\rightarrow+\infty$ where the particle occupation number at a generic lattice site can be equal to $0$ or $1$ only. This considerably simplifies the simulation algorithm (which is based on the stochastic series expansion~\cite{S-99,SS-02,S-92}).

This paper is organized as follows. In Sec. \!II we provide some details on numerical simulations while in Sec. \!III we start our analysis by studying the $\langle n_i\rangle=0$ Mott transition and compare QMC outcomes with the TSS theory. In Sec. \!IV LDA is numerically estimated for the 2D HC BH model and then applied in Sec. \!V in considering the $\langle n_i\rangle=1$ Mott transition. In Sec. \!VI we analyze the superfluid phase by paying particular attention to the scaling properties close to those lattice sites where the effective chemical potential $\mu_{eff}$, which will be defined later, equals approximately $2$. Finally, we conclude in Sec. \!VII.               

\section{Quantum Monte Carlo simulations}
Numerical simulations relied on the directed loop algorithm stemming from the stochastic series expansion method \cite{S-99}: for a generic system with Hamiltonian $H$, its starting point is given by the standard power series expansion of the partition function $Z$, that is
\begin{equation}\label{zed}
Z=Tr\{e^{-\beta H}\}=\sum_{\alpha}\sum_{n=0}^{+\infty}\frac{(-\beta)^n}{n!}\langle\alpha|H^n|\alpha\rangle\ ,
\end{equation}
\noindent being $\{|\alpha\rangle\}$ a basis set. If $H$ can be decomposed as a sum of bond operators $H_{a_i,c_i}$%~\!\footnote{~\!This can be easily done for the HC BH model in any dimension.} 
- where $a_i$ labels the bond and $c_i$ refers to whether the operator is diagonal ($c_i=1$) or not ($c_i=2$) with respect to $\{|\alpha\rangle\}$ -, Eq. (\ref{zed}) can be rewritten as
\begin{equation}
Z=\sum_{\alpha}\sum_{n=0}^{+\infty}\sum_{S_n}\frac{(-\beta)^n}{n!}\langle\alpha|\prod_{i=1}^{n}H_{a_i,c_i}|\alpha\rangle\ ,
\end{equation}
\noindent with $S_n$ standing for a sequence $S_n=[a_1,c_1],\ldots,[a_n,c_n]$.  
\indent We can easily arrange for this setup with the HC BH model. The basis $\{|\alpha\rangle\}$ is chosen to be the set of eigenvectors of the local density operators $n_i$ and this automatically determines which terms in the Hamiltonian are diagonal: contributions with $b_i^\dagger b_j$ have $c=2$ while those written in terms of the $n_i$'s have $c=1$. Moreover, the former are already bond-like while the latter have to be rewritten: as an example,
\begin{equation}
\mu\sum_in_i \rightarrow \mu\sum_{\langle ij\rangle}\Big(\frac{n_i}{f_i}+\frac{n_j}{f_j}\Big)\ ,
\end{equation}
\noindent where the sum on the right-hand side runs on nearest-neighbor sites and where $f_i$ and $f_j$ are the number of links having respectively site $i$ and site $j$ as one end.\\ 
\indent Even though the Taylor expansion above converges \cite{S-92}, statistically relevant contributions are basically provided by configurations where the number of bond operators in Eq.(8) is finite and below an opportune value $N_{\rm tr}$; therefore, truncating the series at order $N_{\rm tr}$ for practical purposes should not entail any significant truncation error, as explained in Sec. IIA of \cite{SS-02}. In determining  $N_{\rm tr}$, we opted for the standard definition, that is we set $N_{\rm tr}=1.5 \, M_{\rm max} \, N_{\rm{bonds}}/T$, where $M_{\rm max}$ is the highest matrix element of the single-bond Hamiltonians and $N_{\rm{bonds}}$ is the number of interacting site pairs. Besides checking that this cutoff was never crossed during the updating process, fluctuations in the order of the series expansion were monitored to control whether the averaged order was consistently less than $N_{\rm tr}$ (with deviations proportional to the square root of the mean value). As proven in \cite{SS-02}, these criteria ensure that the truncation error is negligible compared to the statistical uncertainty stemming from Monte Carlo fluctuations.\\
\indent Exploiting this truncation, the expression for the partition function can be further simplified, i.e.,
\begin{equation}
\label{genzed}
Z=\sum_{\alpha}\sum_{S_{N_{\rm tr}}}\frac{(-\beta)^n(N_{\rm tr}-n)!}{N_{\rm tr}!}\langle\alpha|\prod_{i=1}^{N_{\rm tr}}H_{a_i,c_i}|\alpha\rangle\ ,
\end{equation}
\noindent where $N_{\rm tr}-n$ identity operators have been inserted in all possible ways in the sequence $S_{N_{\rm tr}}$. It is understood that now the index $c_i$ can assume a third value ($c_i=0$) corresponding to the identity itself.\\
\indent In Eq. (\ref{genzed}) the space of configurations have been generalized to be $\{|\alpha\rangle\}\otimes\{S_{N_{\rm tr}}\}$. This can be sampled by means of two kinds of steps: the first type $M_1$ consists of replacing identity operators in the sequence $S_{N_{\rm tr}}$ with diagonal ones (and vice versa), while the second kind $M_2$ is given by exchanging diagonal operators with non-diagonal ones (and vice versa). An exhaustive description of both steps and of how they are performed can be found in Ref. \cite{SS-02}. Let us just recall that, in implementing kind $M_2$, a set of transition probabilities is needed and must be determined by solving so-called directed loop equations. Depending on the parameters of the Hamiltonian, it is possible to select solutions able to reduce the number of bounces,~\!\footnote{~\!This issue is treated in great detail in Sec. IID of \cite{SS-02} where the XXZ model is studied. Since its matrix elements are in one-to-one correspondence with those of the HC BH model, the discussion can be easily adapted to the present case.} that is to cut the amount of moves where the proposed change is rejected. Such solutions are preferred since they shorten the computer time needed to update the configuration. In our simulations, one bounce was allowed within each group of equations at $\mu=0$ and $\mu=2$ while two bounces entered into play when $\mu=-2$.\\   
\indent One MC step is made out of a single step of type $M_1$ followed by a number $N_{\rm{loops}}$ of updates of kind $M_2$. $N_{\rm{loops}}$ is fixed at runtime by imposing that the number of visited vertices is of the order of $N_{\rm tr}$ in a MC step.\\
\indent Runs are performed fixing temperature $T$, chemical potential $\mu$, trap size $l$ and lattice size $L$ with open boundary conditions.
Finite-size effects are avoided by choosing $L$ sufficiently large to obtain $L\to\infty$ data within statistical errors. This condition was fulfilled  taking $L/l\approx 3$ when $\mu=2$ and $\mu=0$ and $L/l\approx 5$ when $\mu=-2$.\\
\indent A standard jackknife was employed to assess errorbars, each bin being the mean of $10^4$ MC step measurements. Typical statistics of our QMC simulations range from $2.5\times10^6$ MC steps for simulations at $\mu=2$ to $7.5\times10^6$ MC steps for simulations at $\mu=0$.

\begin{figure}[tbp]
\vspace*{-0.87cm}
\includegraphics*[width=9cm,height=6.5cm]{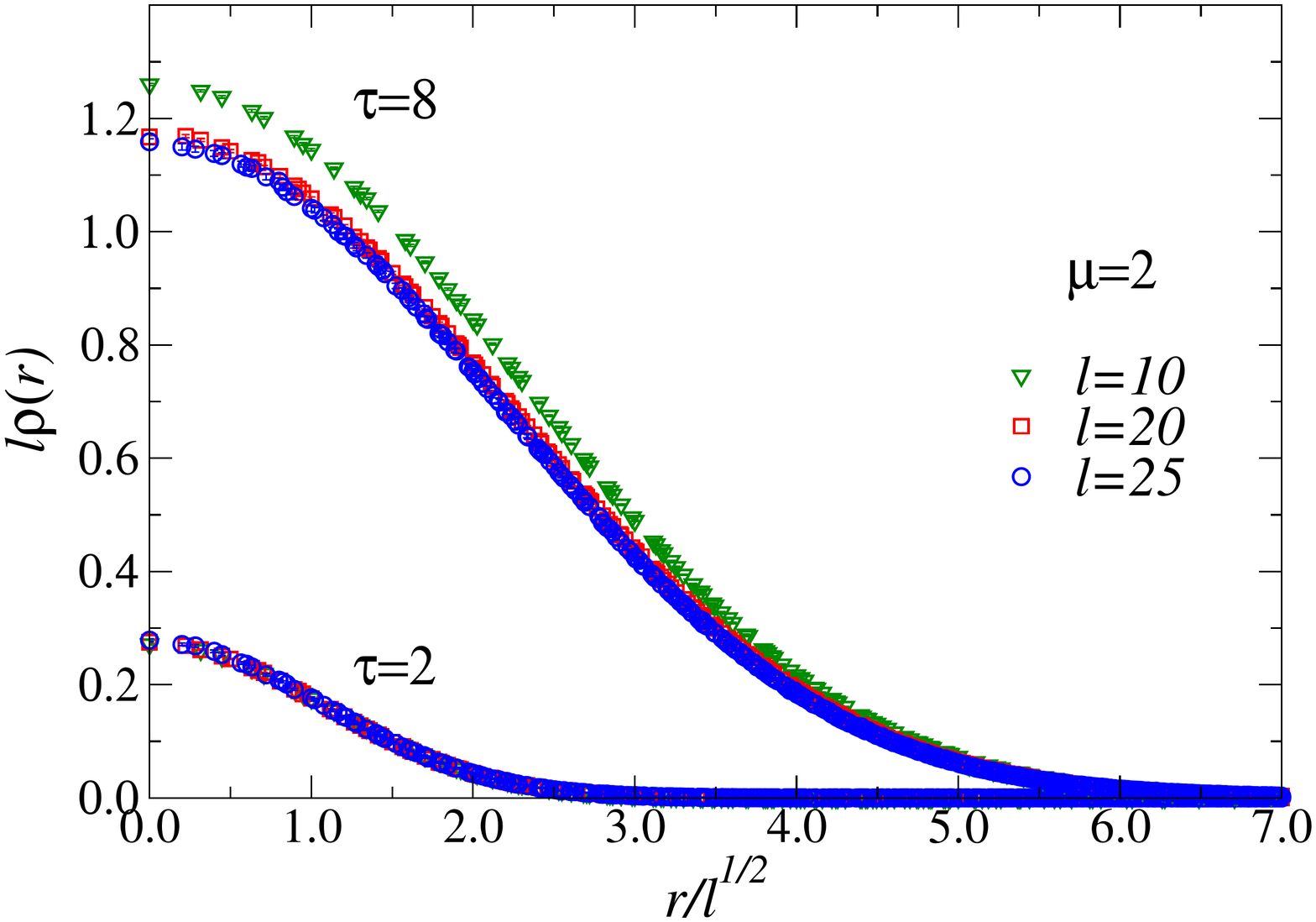}
\vspace*{-0.8cm}
\caption{(Color online) The particle density at $\mu=2$ with $\tau\equiv Tl=8$ and $\tau=2$ for some values of the trap size $l$.}
\label{densmu2}
\end{figure}

\section{The critical point at\ \ \!$\mu=2$}
We now discuss our QMC results for the density and the density-density correlator
at the Mott-insulator to superfluid transition where $\langle n_i\rangle=0$.
In analogy with the singular part of the free-energy density (\ref{freee}), the scaling ansatz for the two above-mentioned observables~\!\footnote{~\!From the conventions introduced after Eq. (\ref{vrp}), it is clear that both $\rho(r)$ and $G(r,r')$ are dimensionless quantities.} read:
\begin{align}
\label{density}
\rho(r) &= l^{-d\theta} {\cal D} ( \bar\mu l^{2\theta} , Tl^{2\theta} , rl^{-\theta} ) \; ,\\  
\label{correlator}
G(r,r') &= l^{-2d\theta} {\cal G} ( \bar\mu l^{2\theta} , Tl^{2\theta} , rl^{-\theta} , r'l^{-\theta} ) \; ,
\end{align}
where the critical exponents for this transition $\nu=1/2$ and $z=2$ have been used. Scaling corrections due to irrelevant perturbations in $l^{-\theta}$ and possible analytic contributions have been neglected.
After setting $d=2$, introducing the scaling coordinates $R=rl^{-\theta}, R'=r'l^{-\theta}$,
and considering the system at criticality (so that $\bar\mu=0$), Eqs. (\ref{density}) and (\ref{correlator}) can be rewritten as
\begin{align}
\label{density2}
l^{2\theta} \rho(r)  &\approx \hat{{\cal D}} (\tau,R) \; , \\
\label{correlation2}
l^{4\theta} G(r,r')  &\approx \hat{{\cal G}} (\tau,R,R') \; ,
\end{align}
being $\tau \equiv Tl^{2\theta}$ the scaling variable that controls the critical behavior of the system.~\!\footnote{~\!From now on, quantities $R$, $R'$ and $\tau$ will always be defined as in this section unless differently specified.}
The meaning of Eqs. (\ref{density2}) and (\ref{correlation2}) should be pretty clear: A given observable rescaled with the proper power of the trap size $l$ equals a universal function depending on $\tau$, $R$, $R'$, etc. Therefore, data obtained via simulations with values of the parameters tuned in such a way to keep the arguments of the function on the right-hand side of Eqs. (\ref{density2}) and (\ref{correlation2}) constant should collapse on a unique curve once that the proper rescaling has been performed. For the Mott-insulator to superfluid transition in the low-density regime, this condition is fulfilled by performing simulations with fixed $Tl$ since $\theta =1/2$.\\
\begin{figure}[tbp]
\vspace*{-0.87cm}
\includegraphics*[width=9cm,height=6.5cm]{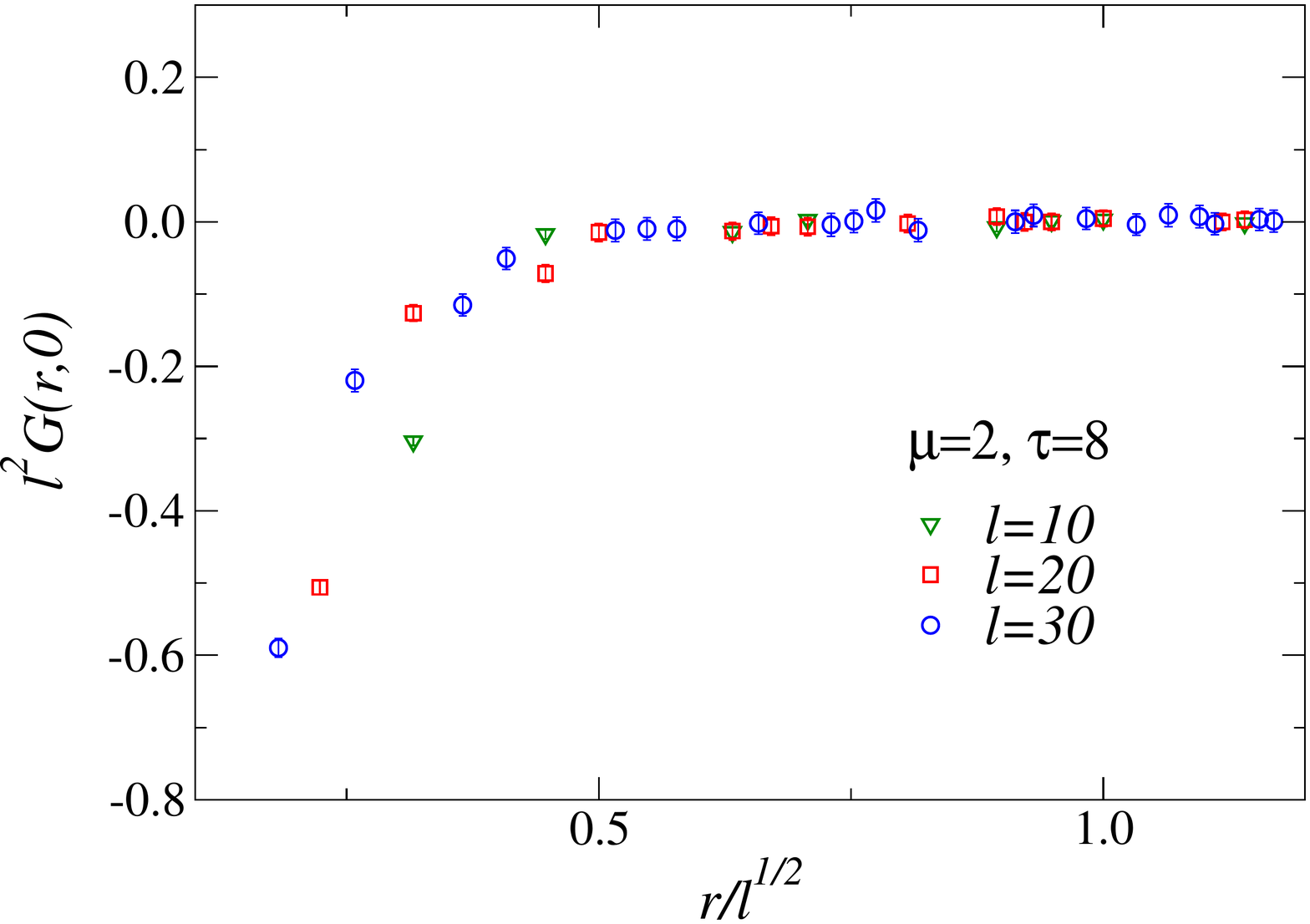}
\vspace*{-0.8cm}
\caption{(Color online) The density-density correlator at $\mu=2$ with fixed $\tau\equiv Tl=8$ for different
values of the trap size $l$.}
\label{corrmu2}
\end{figure}
\indent While in the 1D HC BH model the particle density and the density-density correlator could be treated analytically both at zero and finite temperature (so that numerical outcomes could be compared with their analytical values \cite{CTV}), in the two-dimensional case no exact solution is available. In this study TSS is applied to a 2D-system.\\ 
\indent Figure \ref{densmu2} shows the rescaled particle density. Data in it are divided into two groups corresponding to simulations performed with fixed $\tau=2$ or $\tau=8$. Since $\tau\equiv Tl$ and since the values of $l$ are essentially the same in both groups, sets with $\tau=2$ are generally related to lower temperatures than those with $\tau=8$. While the latter shows scaling corrections at small $l$, it is evident that the former have a more pronounced tendency to collapse on a universal curve. This comes with no surprise since universality is a feature appearing in proximity of a phase transition, which occurs at $T=0$ when working with chemical potential fixed at $\mu=2$ as in the present case.\\ 
%\begin{figure}[t]
%\includegraphics*[width=9cm,height=7cm]{LDA_FIT.pdf}
%\vspace*{-0.8cm}
%\caption{(Color online) Numerical outcomes for $\rho_*({\mu})$ vs. $\mu$ for different values of the lattice extent $L$ and temperature $T$: data were collected in the homogenous system with periodic boundary conditions. The dotted line represents the polynomial fit of the data corresponding to the larger extent.}
%\label{LDA_fit}
%\end{figure}
\indent Figure \ref{corrmu2} contains the rescaled density-density correlator at fixed $\tau=8$ vs.\! $R$. In analogy with the particle density, also for this observable, numerical outcomes after the rescaling prescribed by TSS display a tendency to collapse on a unique curve when increasing $l$, in agreement with the ansatz in Eq. (\ref{correlation2}). Once again, corrections can be noticed only at small values of the trap size.   

\section{Local Density Approximation}
In many statistical systems featuring an external potential $V(r)$ varying with the space position, it is common to approximate the ground-state density at point $r$ with the value that the density assumes in the homogeneous system provided with a constant potential fixed everywhere at the value $V(r)$ that the potential takes at point $r$ itself in the inhomogeneous case. This approximation is called Local Density Approximation (LDA).\\ 
\indent LDA has already been verified to be exact in the 1D HC BH model at zero temperature \cite{CV-10-bh} and it is reasonable to test to which extent it works also in the two-dimensional case at finite $T$. In general, considering a constant potential in Eq. (\ref{bhm}) essentially means to introduce an effective chemical potential $\mu_{\rm eff}(r)$ given by
\begin{equation}\label{effective-potential}
\mu_{\rm eff}(r) \equiv \mu + \frac{r^2}{l^2}\ .
\end{equation} 
Therefore, in analogy with the 1D HC BH model, we assume that the LDA of the 2D trapped system equals
\begin{equation}
\rho_{\rm LDA}(r) = 
\kern-10pt \quad\left\{
\begin{array}{l@{\ \ }l@{\ \ }l}
0 & {\rm for} & \mu_{\rm eff}(r) > 2\ , \\
    \rho_*(\mu_{\rm eff}) &
    {\rm for} & -2 \le \mu_{\rm eff}(r) \le 2\ , \\
1 & {\rm for} & \mu_{\rm eff}(r) < -2\ , \\
\end{array} \right.
\end{equation}
where $\rho_*(\mu)$ is the unknown $T=0$ density of the 2D homogeneous system provided with an effective chemical potential given by Eq. (\ref{effective-potential}).\\
\begin{figure}[t]
\vspace*{-0.87cm}
\includegraphics*[width=9cm,height=7cm]{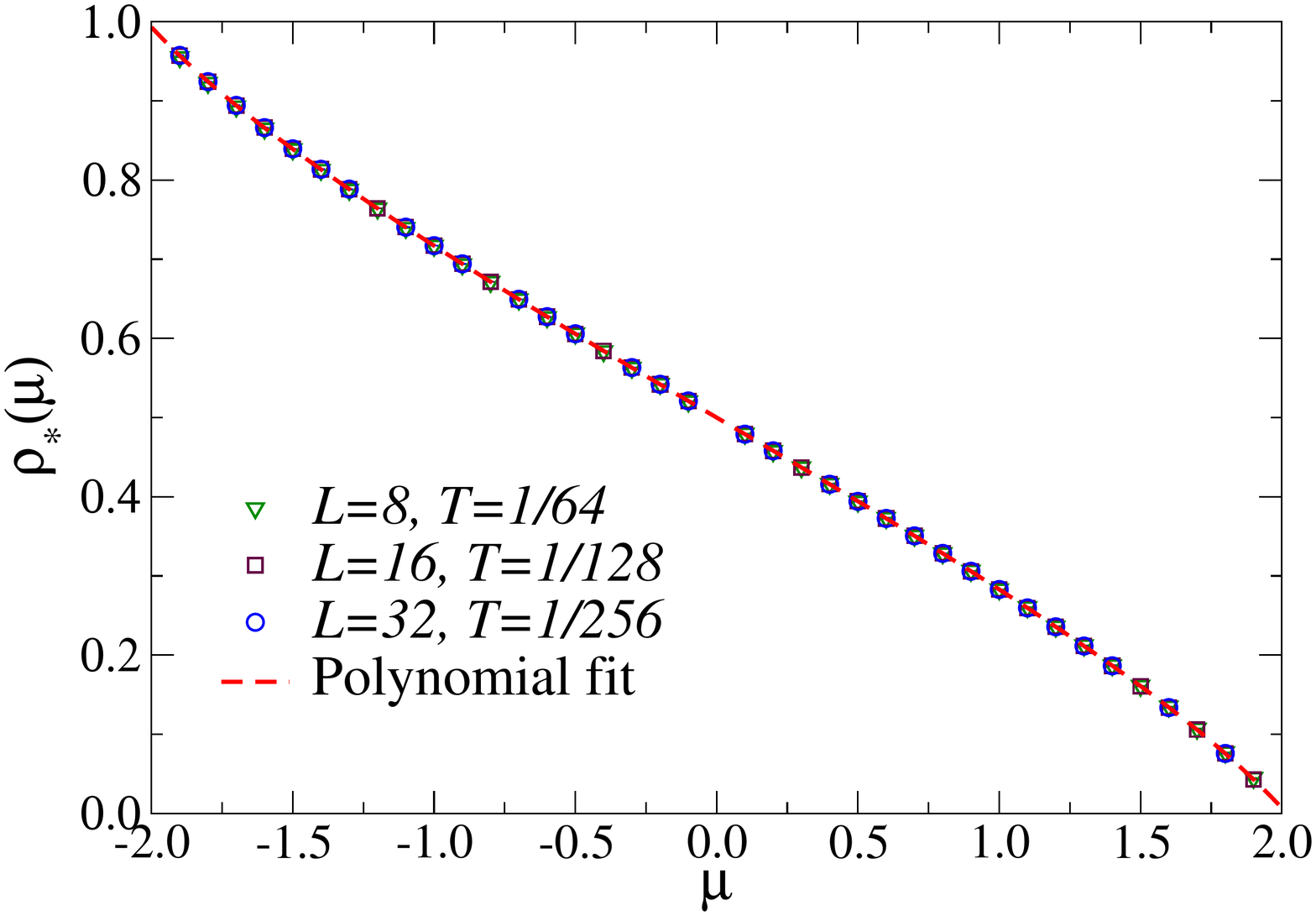}
\caption{(Color online) Numerical outcomes for $\rho_*({\mu})$ vs. $\mu$ for different values of the lattice extent $L$ and temperature $T$. Data were collected in the homogenous system with periodic boundary conditions. The dotted line represents the polynomial fit of the data corresponding to the largest extent.}
\label{LDA_fit}
\end{figure}
\indent In order to obtain an estimate for $\rho_*(\mu)$, we performed simulations of the 2D system without a trap and with periodic boundary conditions~\!\!\footnote{~\!This is expected to reduce finite-size corrections.} in the low-temperature regime for different values of the effective chemical potential. In particular, we employed a set of equally-spaced values covering the range from $-2$ to $+2$.
This setup is easily obtained by setting to zero the trap parameter $v$ in our QMC code and by implementing the specific topology,
all other features of the simulation remaining the same.\\
\indent More specifically, we performed simulations with $L=8$ at $T=1/64$, with $L=16$ at $T=1/128$ and with $L=32$ at $T=1/256$, being $L$ the extent of a square lattice. Following the same criteria of \cite{BBMSTD-02}, we checked that the data were consistent within errorbars so that we could safely assume that the results at $T=1/256$ correspond effectively to the zero-temperature values. Figure \ref{LDA_fit} displays $\rho_*(\mu)$ for the three sets of simulation parameters mentioned above; data basically overlap.\\
\indent Finally we fitted the $T=1/256$ outcomes to a generic polynomial function of degree $n$:
\begin{equation}\label{fit-function}
\rho_*(\mu) = \sum_{i=0}^n c_i \, \mu^i \ , 
\end{equation}
where $n$ was chosen by truncating this Taylor expansion when the $\chi^2$ of the fit stabilized. This was the case with $n=7$ (the reduced $\chi^2$ being approximately $1.5$) though truncations at higher order were also considered without showing meaningful deviations. As expected on theoretical grounds, it turned out that the constant term $c_0$ read $1/2$ (within $10^{-7}$) while even terms were negligible; thus, the only non-trivial contributions are given by the odd powers for which the following estimates were obtained:  
\begin{eqnarray}\label{fit-coefficients}
&&c_1=-0.20779(1) \;\;\; ,\ c_3=-0.01323(1) \;\;\;\nonumber\\
&&c_5=+0.00441(1) \;\;\; ,\ c_7=-0.00093(1) \; .
\end{eqnarray}
\indent The function in Eq. (\ref{fit-function}) with $n=7$ and coefficients as given above is plotted in Fig. \ref{LDA_fit} and was used for the data analysis reported in the following sections.
%We will see that also for the 2D model the convergence to its LDA is fully satisfied in the large-$l$ limit, so that
%only the behavior of the difference $\Delta\rho\equiv\rho - \rho_{\rm LDA}$ will need a theoretical explanation, which will
%be accounted for by TSS.
\begin{figure}[tbp]
\vspace*{-0.87cm}
\includegraphics*[width=9cm,height=7cm]{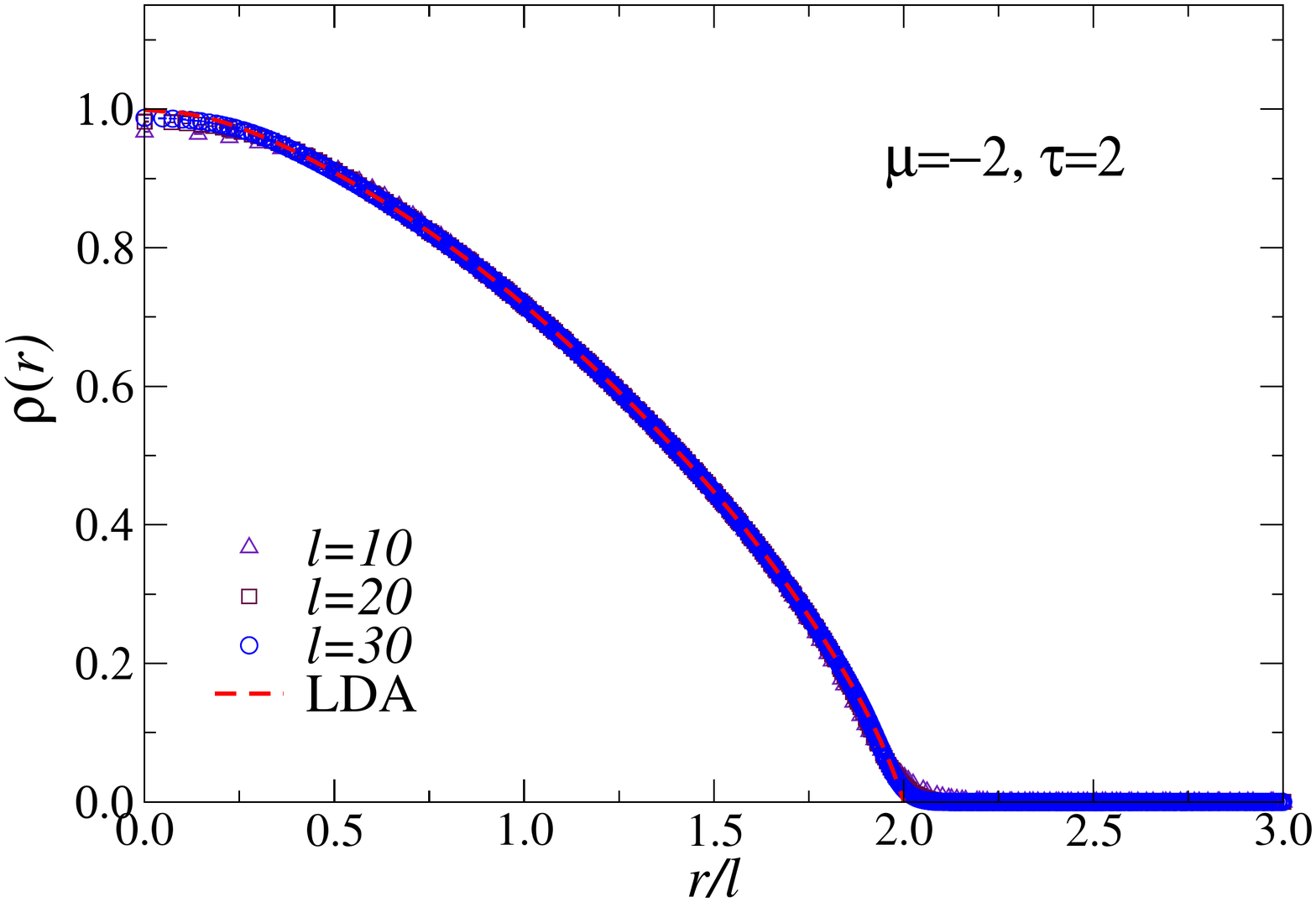}
\includegraphics*[width=9cm,height=7cm]{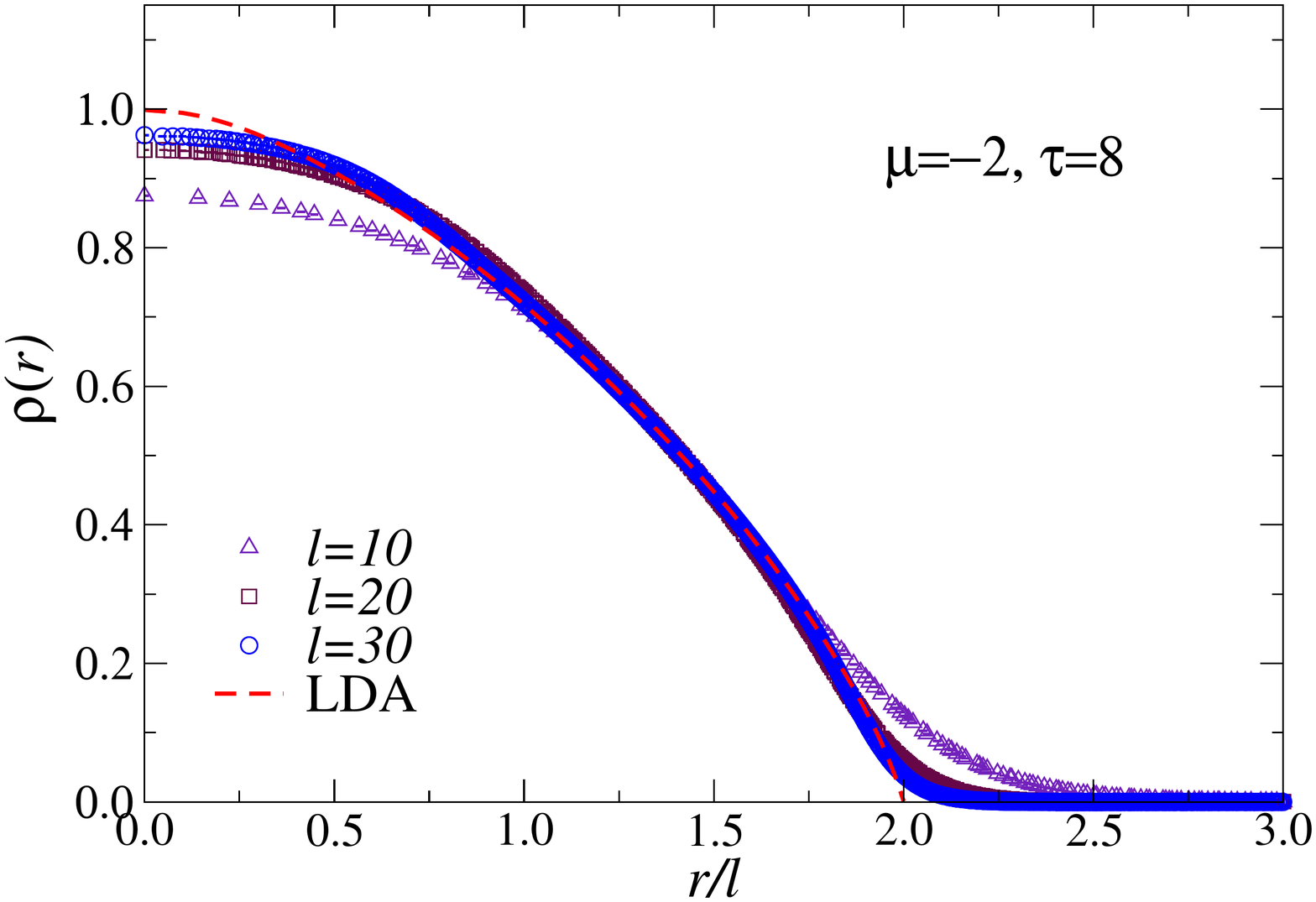}
\includegraphics*[width=9cm,height=7cm]{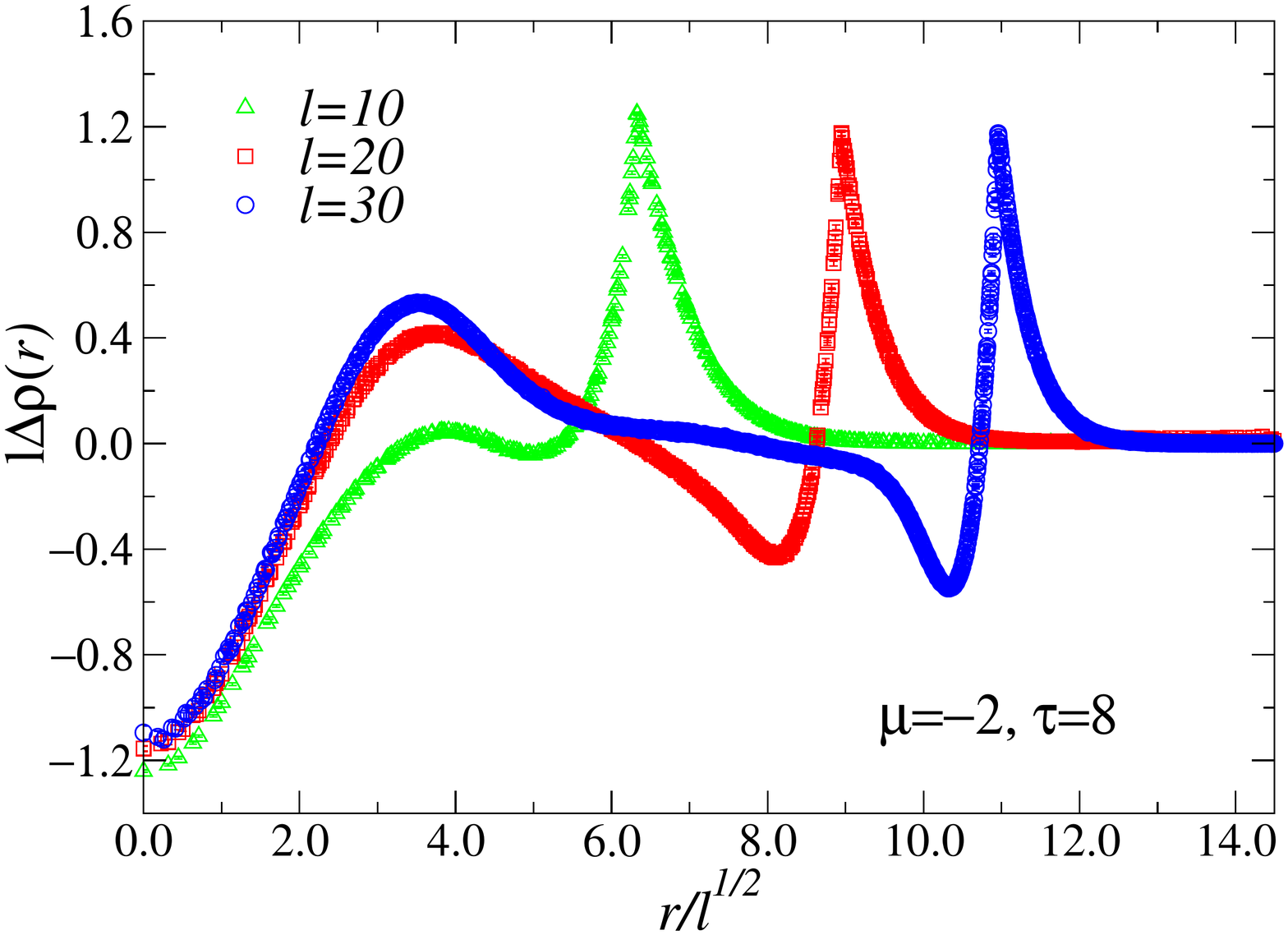}
\vspace*{-0.8cm}
\caption{(Color online) The particle density at $\mu=-2$ with fixed $\tau\equiv Tl=2$ (top) and $\tau=8$ (middle) for different
values of the trap size $l$ and the scaling of the subtracted particle density at $\mu=-2$ with $\tau=8$ (bottom). The dotted line in the first two plots represents the numerical estimates of the LDA.}
\label{densmu-2}
\end{figure}
\section{$n=1$\ \ \! Mott transition}
The invariance under the particle-hole exchange entails a similar behavior of the homogeneous HC BH model at the transitions with $\mu=2$ and $\mu=-2$.
However, the trap-size scaling behavior at the $\langle n_i\rangle=1$ transition is expected to be different than in the vacuum-to-superfluid one because the particle-hole symmetry does not hold for a trapped system.\\
\indent Studies of the trapped 1D HC BH model at zero and finite temperature \cite{CV-10-bh,CTV} have revealed how, at the superfluid to Mott transition with non-zero filling, the particle density approaches its Local Density Approximation (LDA) in the large-$l$ limit. In analogy with the one-dimensional case, we thus expect this observable to be given by an expression like
\begin{eqnarray}
\label{LDA}
\rho(r) &=& \rho_{\rm LDA}(rl^{-1}) + l^{-2\theta} \hat{{\cal D}} (Tl^{\theta z},rl^{-\theta})\ =\nonumber\\ 
&=& \rho_{\rm LDA}(rl^{-1}) + l^{-1} \hat{{\cal D}} (\tau,R) \; ,
\end{eqnarray}
since $\theta=1/2$ and $z=2$ again; irrelevant corrections in $l^{-\theta}$ have been neglected once more. \!Therefore, the scaling quantity should not be the density itself but rather the difference
\begin{equation}
\label{drho}
\Delta\rho(r)\equiv\rho(r)-\rho_{\rm LDA}(rl^{-1})\ .
\end{equation} 

\begin{figure}[t]
\vspace*{-0.87cm}
\includegraphics*[width=9cm,height=7cm]{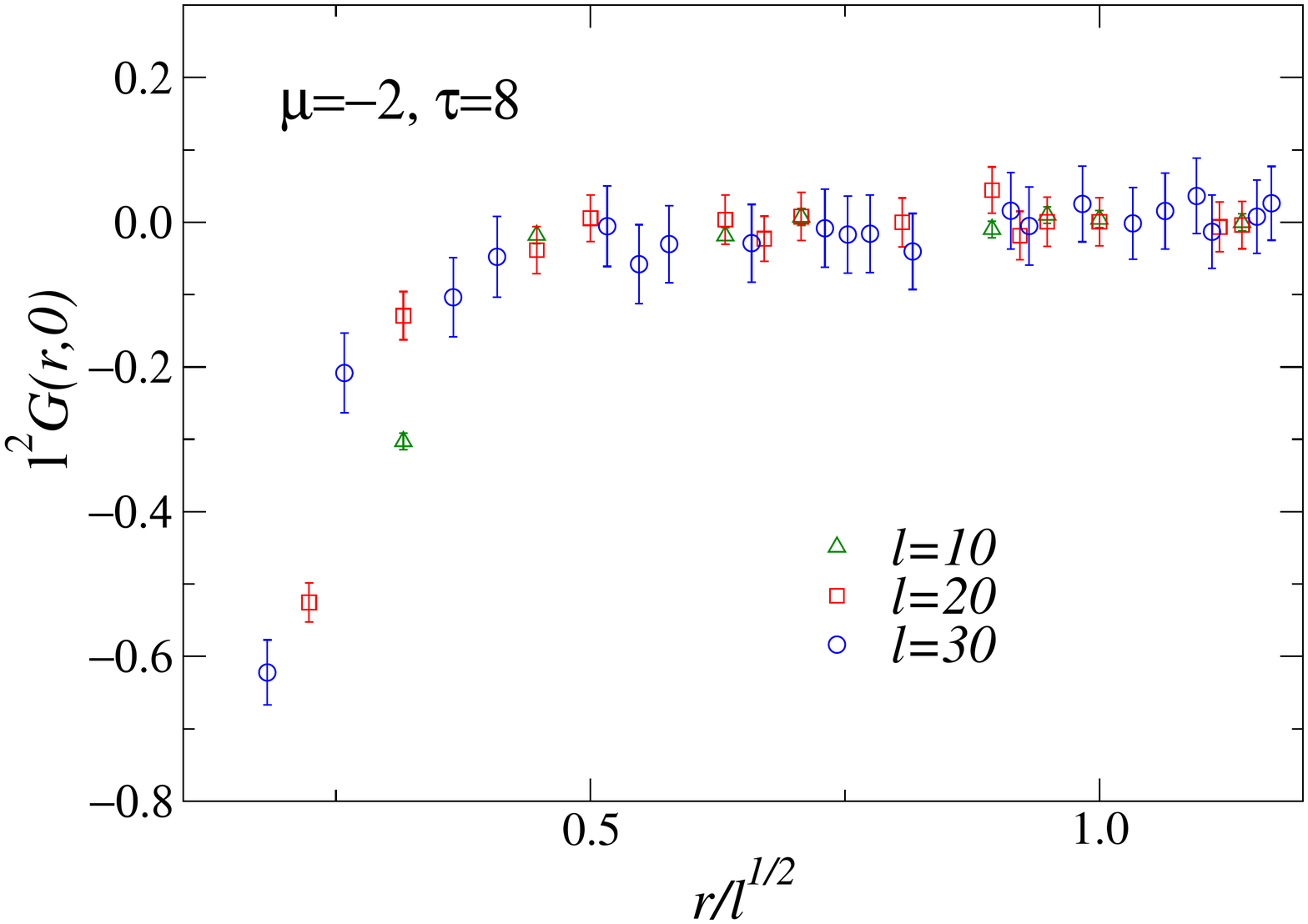}
\vspace*{-0.8cm}
\caption{(Color online) Scaling of the density-density correlator at $\mu=-2$ with fixed $\tau=8$.}
\label{densmu-2s}
\end{figure}
\indent Figure \ref{densmu-2} shows how the particle density converges to the LDA in the two-dimensional model. Since $\tau\equiv Tl$ as in Sec. III, once again data sets with $\tau=2$ correspond to temperatures lower than those of the sets with $\tau=8$, given the common values of the trap size. Besides improving with increasing $l$ in agreement with Eq. (\ref{LDA}), the convergence to the LDA is better at small $T$, as clear from comparing the upper and middle part of Fig. \ref{densmu-2}, since LDA itself is approached at $T\rightarrow0$.\\
\indent The lower part of Fig. \!\ref{densmu-2} illustrates the behavior of $\Delta\rho(r)$ at $\tau=8$ after the rescaling suggested by Eq. (\ref{LDA}) has been performed: a tendency to collapse on a universal curve is evident in a region close to the origin while some transition-like peaks appear at a distance $r\approx2l$ from the center, thus drifting with increasing $l$. Therefore, in the $l\!\!\rightarrow\!\!+\infty$ limit, these peaks disappear and only the region with the universal curve is left, hence confirming TSS predictions in Eq. (\ref{LDA}). The deviations from the expected scaling at finite $l$ will be treated in Sec. VI.\\   
\begin{figure}[t]
\vspace*{-0.87cm}
\includegraphics*[width=9cm,height=7cm]{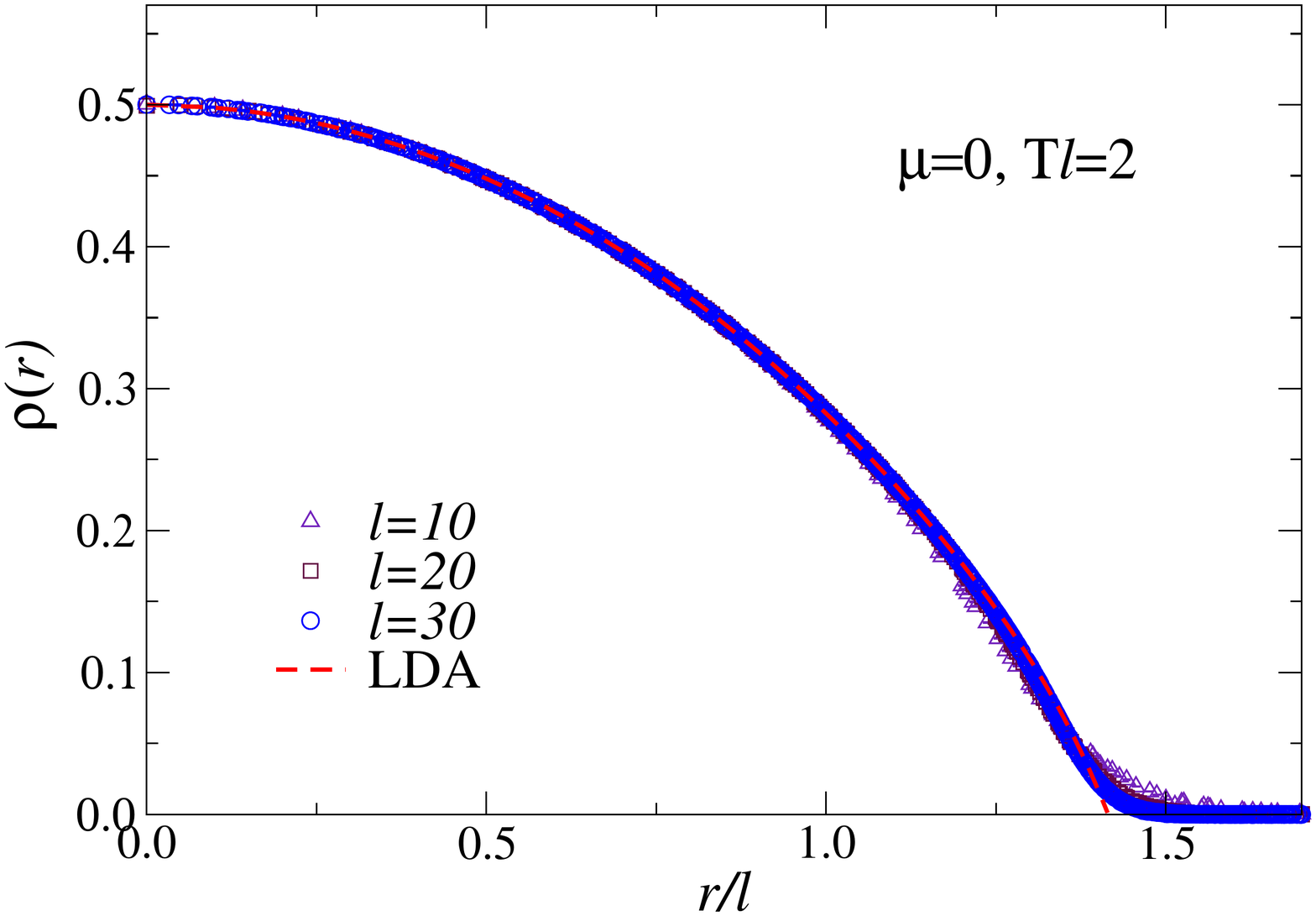}
\includegraphics*[width=9cm,height=7cm]{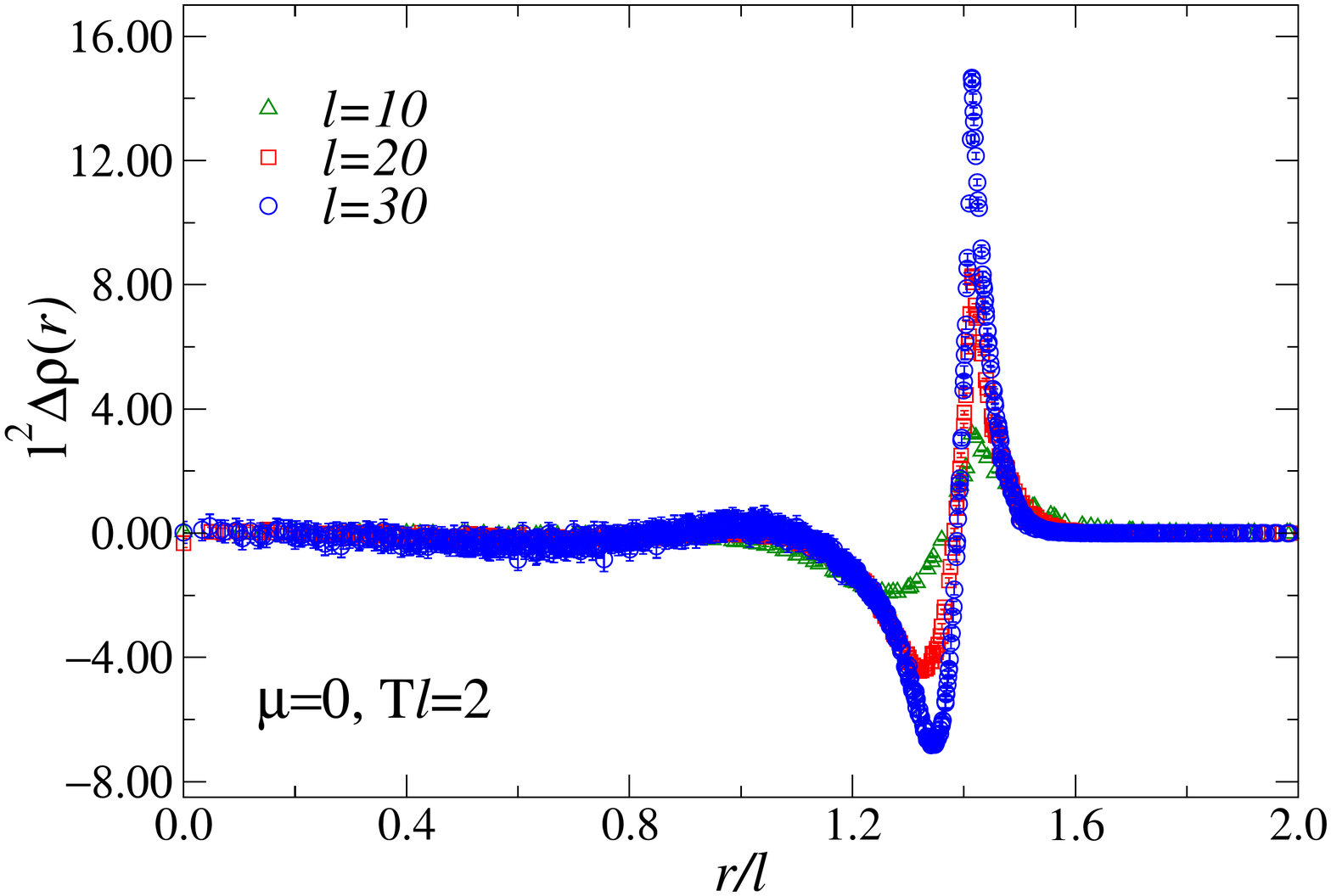}
\vspace*{-0.8cm}
\caption{(Color online) The particle density vs. $r/l$ (top) and the rescaled subtracted particle density (bottom) at $\mu=0$ with fixed $\tau\equiv Tl=2$.}
\label{densmu0noresc}
\end{figure}
\indent As for the density-density correlation function, scaling expectations, again given by Eq. \!\!\!(\ref{correlation2}), are also nicely confirmed, as depicted in Fig. \!\ref{densmu-2s}. At $\tau=8$, besides scaling corrections at small trap size, no strong deviations from a universal curve could be detected within errorbars.  In the 1D HC BH a striking result was the universality for the correlator between Mott-to-superfluid transitions at $\langle n_i\rangle=0$ and at $\langle n_i\rangle=1$. A closer inspection at Figs.\! \ref{corrmu2} and \ref{densmu-2s} reveals how this feature seems to hold in the 2D case as well.\\
\indent In the 1D HC BH at zero temperature \cite{CV-10-bh}, a peculiar characteristic of scaling quantities like the subtracted particle density or the density-density correlator was the appearance of modulations depending on the trap size $l$. It was shown that, at fixed $\mu<1$, there are values of the trap size $l$, whose number increases with $l$ itself, for which the energy gap $\Delta E$ between the ground state and the first excited state vanishes. This repeated level crossing leads to modify the scaling ansatz like Eqs. (\ref{density2}) and (\ref{correlation2}) in order to include a further dependence of the universal function on a phase $\phi$ related to the difference between values of $l$ corresponding to zeros of $\Delta E$. At finite temperature, it is expected that this phenomenon plays a lesser and lesser role with increasing $T$ since thermal fluctuations should prevail on quantum effects. This has already been checked in the 1D BH model \cite{CTV} and is confirmed also in two dimensions as Figs. \ref{densmu-2} and \ref{densmu-2s} reveal.~\!\footnote{~\!Recalling the phase diagram, it has to be borne in mind that the condition $\mu<1$ in 1D corresponds to $\mu<2$ in 2D.}         

\section{The spatial region where\ \! $\mu_{\rm eff}=2$}
We are now going to study the 2D BH model at $\mu=0$. Since this corresponds to the deep interior of the superfluid phase, no transition is expected to be monitored. However, a closer look at Eq. (\ref{bhm}) suggests to group the last two terms of the BH Hamiltonian, thus giving rise to the effective chemical potential $\mu_{\rm eff}(r)$ already introduced in Eq. (\ref{effective-potential}).\\ 
\begin{figure}[t]
\vspace*{-0.87cm}
\includegraphics*[width=9cm,height=7cm]{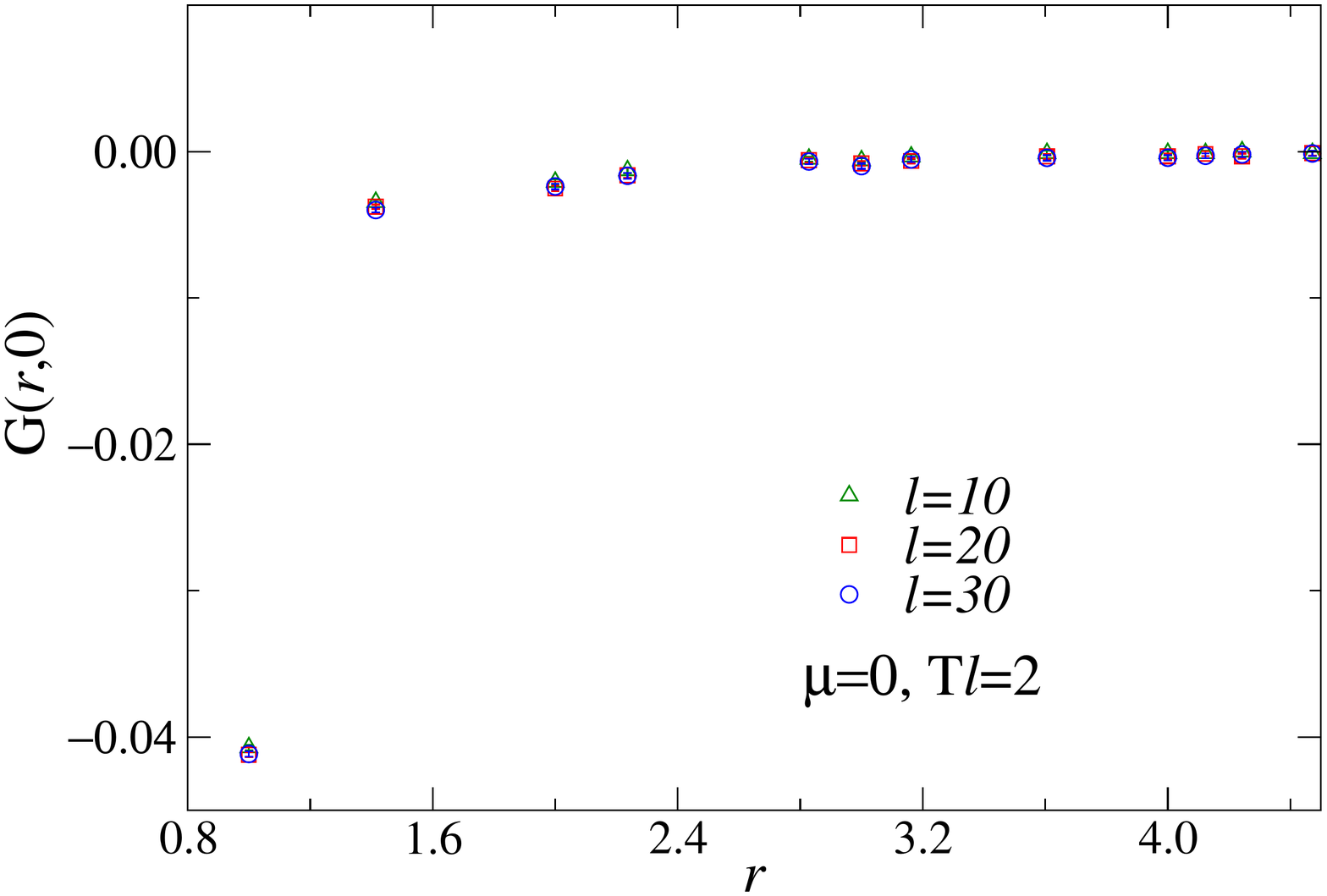}
\vspace*{-0.8cm}
\caption{(Color online) The density-density correlator at $\mu=0$ with fixed $\tau\equiv Tl=2$.}
\label{corrmu0noresc}
\end{figure}
\begin{figure}[t]
\vspace*{-0.87cm}
\includegraphics*[width=9cm,height=7cm]{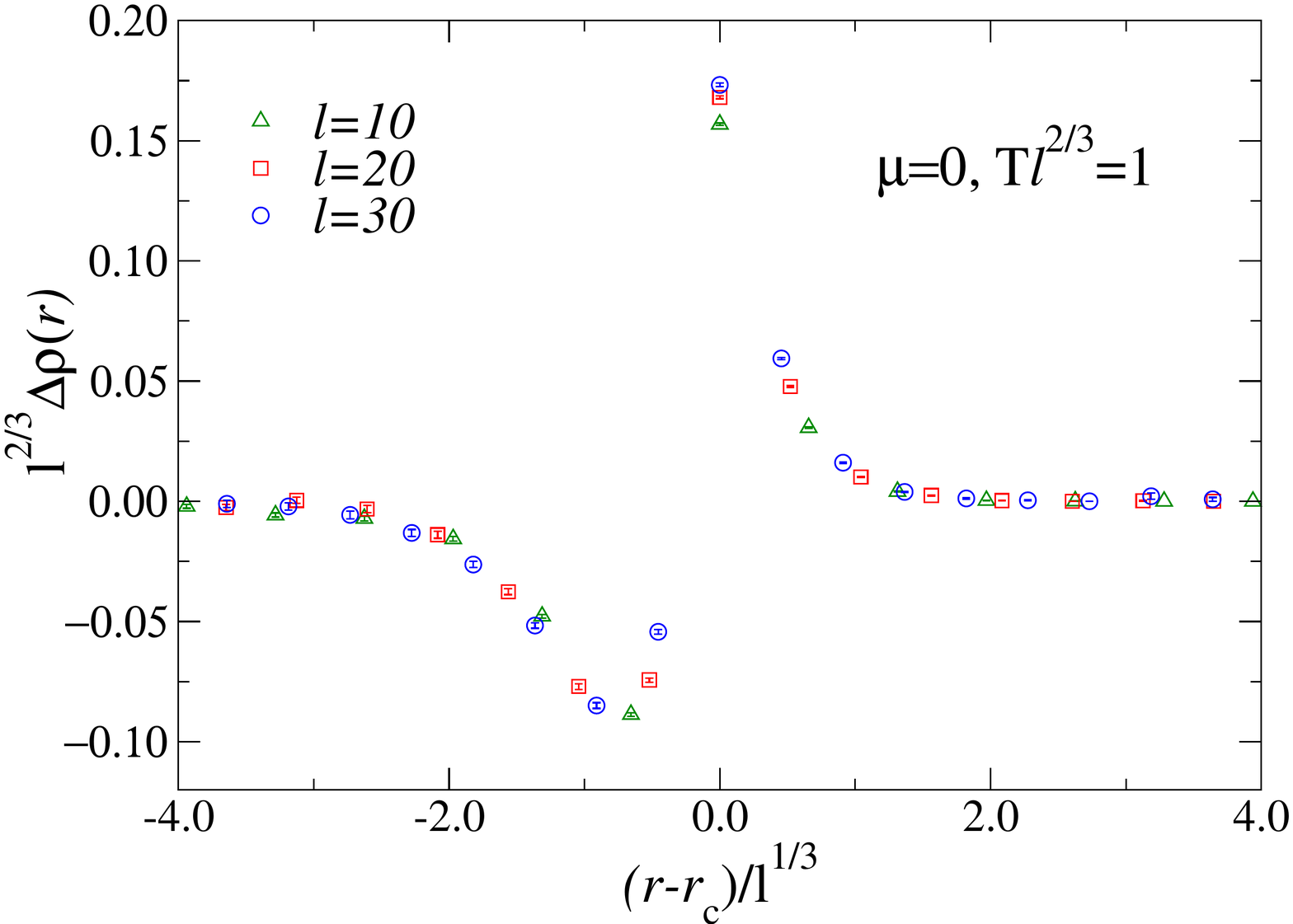}
\includegraphics*[width=9cm,height=7cm]{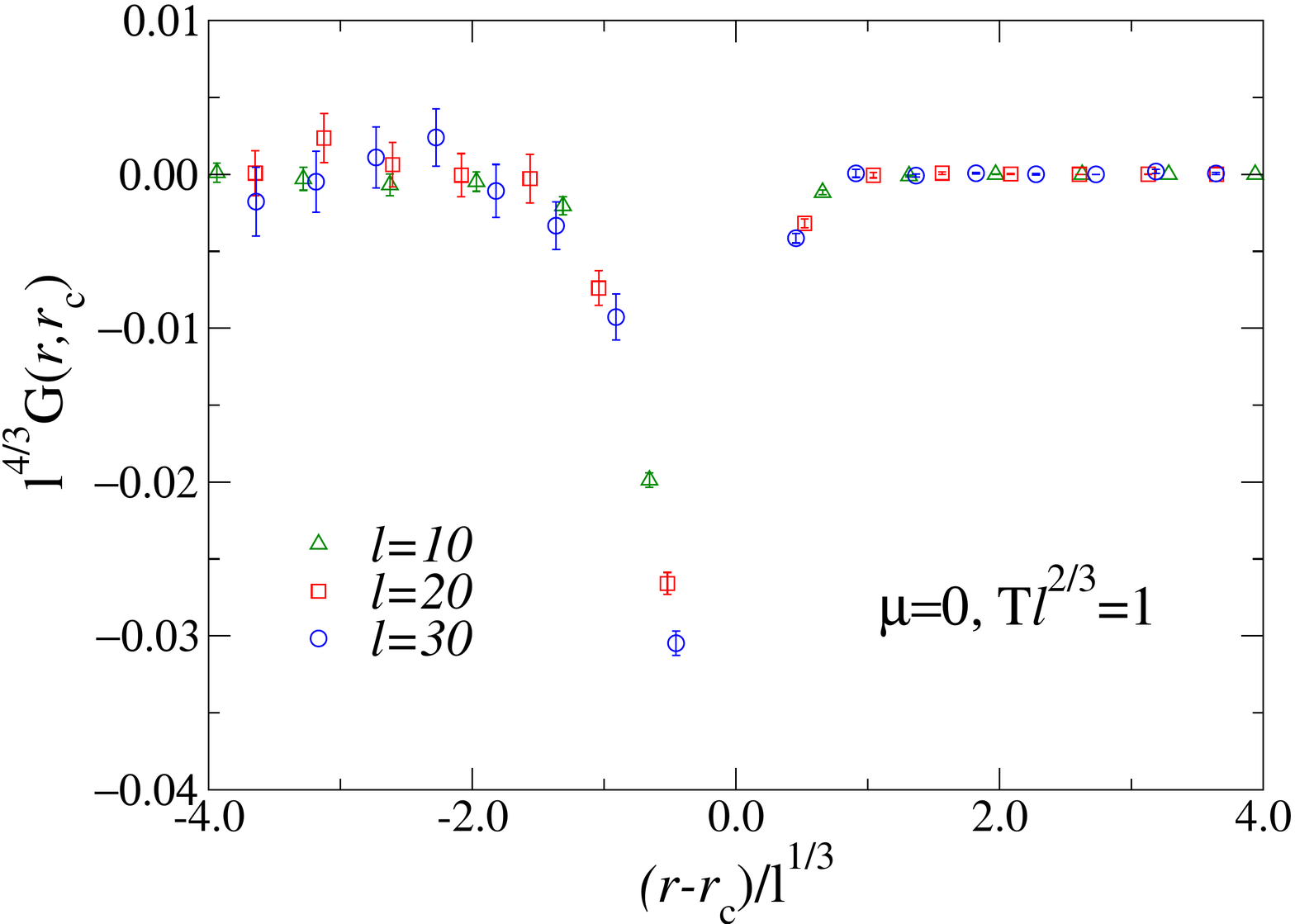}
\vspace*{-0.8cm}
\caption{(Color online) Scaling of the subtracted particle density (top) and the density-density correlator (bottom) at $\mu=0$ with fixed $Tl^{2/3}=1$ around distance $r_c$ where $\mu_{\rm eff}=2$.}
\label{denscorrmu0}
\end{figure}
\indent After this step, it turns out that, in those regions where $\mu_{\rm eff}(r)\approx 2$, the Hamiltonian is effectively given by that of the homogeneous system at criticality;~\!\footnote{~\!The number of dimensions is understood to be $d=2$, otherwise a critical regime would appear in different sectors of the lattice.} therefore, we might expect some sort of phase transition and universal behavior in these regions even if $\mu$ is not set to a critical value.\\
\indent This seems to be indeed confirmed by the behavior of the particle density portrayed in Fig. \ref{densmu0noresc}. If $\Delta\rho(r)$ defined in Eq. (\ref{drho}) is rescaled with $l$ and plotted versus $r/l$ as suggested by Eq. (\ref{LDA}),~\!\footnote{~\!Recall that $z=1$ in the superfluid phase while $\theta=1$ for smooth modes according to an ansatz verified in Ref. \cite{CV-10-bh}.} no substantial difference with the LDA is observed except for some transition-like peaks at ``critical" distance $r_c=\sqrt{2}l$, corresponding to the distance from the center obtained from Eq. (\ref{effective-potential}) after setting $\mu_{\rm eff}(r)=2$ and $\mu=0$.\\
\indent We have already encountered such a situation in Sec. V. Indeed, by setting $\mu_{\rm eff}(r)=2$ and $\mu=-2$ again in Eq. (\ref{effective-potential}), this results in $r=2l$, corresponding to the distance at which peaks were observed in the $\langle n_i\rangle=1$ Mott-insulator phase (see lower part of Fig. \!\ref{densmu-2}).\\
\indent In order to study the scaling in such spacial sectors of the system, an expansion of $\mu_{\rm eff}(r)$ around $r_c$ is needed, i.e.,
\begin{equation}
\mu_{\rm eff}(r) = 2 + p(2-\mu )^{1-1/p} \, \frac{r-r_c}{l} + O[(r-r_c)^2] \ .
\end{equation}
As pointed out in Ref. \cite{CTV}, the length scale $\xi$ of these critical modes should behave like $\xi\approx l^{\sigma}$, $\sigma$ being the exponent associated to a linear potential. In other words, $\sigma=1/3$ \cite{CV-10-bh} and, again neglecting irrelevant contributions in $l^{-\theta}$, scaling should read  
\begin{align}
\label{dr2}
l^{2/3} \Delta\rho(r) &\approx \hat{{\cal D}} (\tau,R) \; , \\
\label{co2}
l^{4/3} G(r,r_c) &\approx \hat{{\cal G}} (\tau,R) \; ,
\end{align}
where $R$ and $\tau$ correspond now to $R=(r-r_c)/l^{1/3}$ and $\tau\equiv Tl^{2/3}$. In deriving the latter exponent the value $z=2$ has been employed since we expect critical modes close to $r_c$ to be controlled by the same dynamical exponent as in the regular Mott-to-superfluid transitions. The exponents in Eqs. (\ref{dr2}) and (\ref{co2}) do not coincide with those in Fig. \ref{densmu0noresc}, where also $T$ and $l$ are not properly tuned, and this explains why data sets do not collapse on a unique curve around $r_c=\sqrt{2}l$. Because of the ``improper" values of the trap size and the temperature, also the particle-particle correlator $G(r,0)$ in Fig. \ref{corrmu0noresc} does not show any particular scaling but simply vanishes after a few lattice spacings.\\
\indent Figure \ref{denscorrmu0} shows the behavior of $\Delta\rho(r)$ and $G(r,r_c)$ vs. $(r-r_c)/l^{1/3}$ at $Tl^{2/3}=1$ after these observables have been rescaled according to Eqs. (\ref{dr2}) and (\ref{co2}). The collapsing of both quantities on a single curve proves the foreseen scaling clearly right.
%\vspace*{0.7cm}

\section{Conclusions}
We have studied the scaling properties of the two-dimensional Bose-Hubbard model at finite temperature in the presence of a trapping potential at the Mott-insulator to supefluid transitions. The interest in this system and its properties is not only theoretical given that 2D experiments involving cold atoms in optical lattices are currently being carried out~\cite{SPP-07,JCLPPS-10,SPP-08}.\\
\indent Particular attention has been paid to the particle density and the density-density correlator, both computed by means of QMC simulations in the hard-core limit $U\to+\infty$. The latter choice is motivated by the fact that the on-site coupling $U$ should play no role in determining the main properties of the system at criticality. A comparison between numerical outcomes and TSS ansatz has subsequently been performed, revealing that theoretical expectations are well-motivated.\\
\indent An interesting feature arising from our study is how LDA compares to the particle density $\rho(r)$ at finite temperature in two dimensions. As in the one-dimensional case \cite{CTV}, it turns out that also in the 2D HC BH model $\rho(r)$ rapidly converges to the LDA when increasing $l$ even at $T>0$, regardless the value of the chemical potential $\mu$. Even if LDA turned out to be broken in frustrated systems \cite{LM-11} and some shortcomings were proven also in describing the $T=0$ phase diagram of the BH model in two dimensions \cite{MDKKTT-11}, this work shows that LDA is capable of describing at least some properties of the 2D BH model quite nicely, also outside its theoretical range of application.

\acknowledgements \vspace*{-0.08cm} We warmly thank E. Vicari for his suggestions as well as for a careful reading of a first draft of this manuscript. The QMC simulations were performed at the Istituto Nazionale di Fisica Nucleare (INFN) Pisa GRID DATA center, using also the cluster CSN4.

\end{document}